\documentstyle[aps,pra,psfig,twocolumn]{revtex}

\begin{document}
\title{Quantum non ideal measurements}
\author{Jean-Michel Courty\thanks{%
courty@spectro.jussieu.fr}, Francesca Grassia \thanks{%
grassia@spectro.jussieu.fr} and Serge Reynaud\thanks{%
reynaud@spectro.jussieu.fr}}
\address{Laboratoire Kastler Brossel\thanks{%
Unit\'{e} mixte de recherche de l'Universit\'{e} Pierre et Marie Curie, de%
\newline
l'Ecole Normale Sup\'{e}rieure et du Centre National de la Recherche\newline
Scientifique}\thanks{%
website: www.spectro.jussieu.fr/Mesure}, Case 74, 4 place Jussieu,\\
F-75252 Paris Cedex 05, France}
\maketitle

\begin{abstract}
We analyse non ideal quantum measurements described as scattering processes providing an
estimator of the measured quantity. The sensitivity is expressed as an
equivalent input noise. We address the Von Neumann problem of chained measurements
and show the crucial role of preamplification.
\end{abstract}

\section{Introduction}

Active systems are fundamental elements in high precision measurements in
particular measurement on quantum systems. Amplifiers are used either for
amplifying the signal to a macroscopic level or to make the system work
around its optimal working point with the help of feedback loops. The
analysis of sensitivity limits in these devices rises many questions related
to fundamental processes as well as experimental constraints. How does the
active feedback act on the quantum system? How does the coupling with the
environment influence the sensitivity of the measurement? How are these
process related to the fluctuation dissipation theorem? How do the
experimental constraints interplay with the fundamental limitations of the
sensitivity?

The aim of the present paper is to address the question of measurement on
quantum systems with real measurement devices. We address this problem with
quantum network theory. This approach provides a rigorous thermodynamical
framework able to withstand the constraints of a quantum analysis of the
measurement. In the same time, it makes possible a realistic description of
real measurement devices. \ Thermodynamic and quantum fluctuations are
treated in the same footing. The measurement process is described as a
scattering process allowing for a modular analysis of real quantum systems.
Active systems such as the linear amplifier or the ideal operational
amplifier are described in this framework. Here, the approach will be
illustrated by analyzing the sensitivity of a cold damped capacitive
accelerometer developed for fundamental physics applications in space \cite
{Bernard91,Touboul92,Willemenot97}.

We first present the analysis of a quantum measurement with a passive
systems in term of quantum networks. Then, we use this approach to present
the quantum analysis of measurements with an operational amplifier working
in the ideal limit of infinite gain, infinite input impedance and null
output impedance and we illustrate the theoretical framework with the
example of a cold damped accelerometer. We conclude with an analysis of
chained measurements

\section{Quantum Networks}

The measurement of a physical quantity or signal corresponds to the
obtention of a result or readout from a measureing device. In general, this
is obtained after a few stages. Many times, the signal is transduced into an
electrical signal which is amplified. After further processing, the
resulting voltage drives a displaying or a storing device. The final result
is thus obtained after a succession of stages.

As far as measurement at the quantum level is concerned, the principle is
the same: a chain of processes leads from the quantum system to the readout
of the meter. As a first insight into a quantum analysis of measurement, we
will consider the detection of an electromagnetic wave with an antenna and
first restrict our attention to the transduction of the wave into an
electrical signal.

\vspace{4mm}

\begin{figure}[htb]

\centerline{\psfig{figure=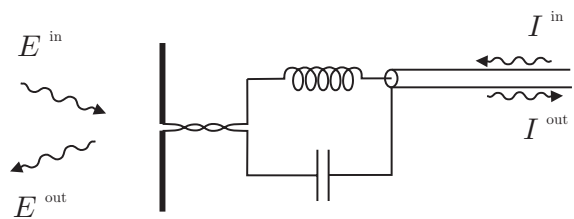,width=7.5 cm}}
\vspace{6mm}

\caption{Model for the measurement of an electromagnetic field. An incoming
electromagnetic wave $E^{\rm in}$ is detected by an antenna coupled to an
electric resonator. The detection is provided by the current $I^{\rm out}$
leaving the coaxial line.The coupling of the measuring device with the
environment by the antenna is also at the origin of dissipation due to the
emission of a field $E^{\rm out}$. The coupling with the line is at the
origin of current fluctuations $I^{\rm in}$ incoming by the line }
\label{Antenne}
\end{figure}

\newpage

As sketched on figure \ref{Antenne}, a detecting antenna is coupled with an
electrical resonator a coaxial line is used to extract an electrical signal
from this system. The further steps as for instance amplification will be
considered in the next section.

The the antenna is coupled to the electromagnetic field though one mode
obtained as a linear combination of plane waves or spherical waves and
corresponding to the radiation diagram of the antenna. In this mode two
counterpropagating fields can be separated an incoming field $E^{{\rm in}} $
and an outgoing field $E^{{\rm out}}$. The incoming electromagnetic wave $E^{%
{\rm in}}$ puts into motion the electrons in the antenna and produces a
current flow in the circuit. As a consequence of this current, a signal $I^{%
{\rm out}}$ with informations on the detected radiation is sent into the
coaxial line. This electrical signal may then be used to feed a further
element, for instance an amplifier. In addition, any incoming signal $I^{%
{\rm in}}$ arriving in the coaxial line also creates a current in the
detector and can perturb the measurement. For example, if the line is at
thermal equilibrium, it corresponds to the Johnson Nyquist noise. Finally,
energy and information can be lost because of the emission of radiation by
the antenna.

These phenomena, present in every measurement can be synthetized in the
following way. In the frequency domain the output fields can be written as
linear combinations of input fields. Throughout the paper, we will consider
that a function $f$ is defined in the time domain (notation $f\left(
t\right) $) or in the frequency domain (Kubo's notation $f\left[ \omega %
\right] $) and that these two representations are related trough the Fourier
transform with the convention of quantum mechanics 
\begin{equation}
f\left( t\right) =\int \frac{d\omega }{2\pi }f\left[ \omega \right]
e^{-i\omega t}
\end{equation}
The electronics convention may be recovered by substituting $j$ to $-i$.

The{\it \ detection and measurement }is then described as a scattering
process: 
\begin{equation}
\left( 
\begin{array}{c}
I^{{\rm out}} \\ 
E^{{\rm out}}
\end{array}
\right) =\left( 
\begin{array}{cc}
\alpha & \beta \\ 
\gamma & \delta
\end{array}
\right) \left( 
\begin{array}{c}
I^{{\rm in}} \\ 
E^{{\rm in}}
\end{array}
\right)
\end{equation}
The measurement of the field $E^{{\rm in}}$ with the current $I^{{\rm out}}$
can be described with an estimator$\hat{E}^{{\rm in}}$%
\begin{equation}
\hat{E}^{{\rm in}}=\frac{1}{\beta }I^{{\rm in}}=E^{{\rm in}}+\frac{\alpha }{%
\beta }I^{{\rm in}}
\end{equation}
the incoming current fluctuations correspond to a noise in the measurement.
It is also possible to express the output field $E^{{\rm out}}$%
\begin{equation}
E^{{\rm out}}=\gamma I^{{\rm in}}+\delta E^{{\rm in}}
\end{equation}
$\gamma I^{{\rm in}}$ coreponds to the back action noise on the measured
device. This kind of description of a measurement can be generalized to more
complex systems in the framework of quantum network theory.

In this approach, the role of the antenna and the line are to couple the
measurement device with the outside world for inputs as well as for outputs.
This coupling is associated to {\it dissipation and fluctuations }which are
thus inavoidable in measurement processes{\it . }Dissipation corresponds to
the loss of energy by the resonator as emitted radiation by the antenna or
as an electrical signal by the line. Fluctuations corresponds to the
detection of the thermal and quantum fluctuations of the input fields. Since
Einstein and its study of the viscous damping of mechanical systems \cite
{Einstein05} the link between fluctuations and dissipation has been widely
studied. For example with Johnson-Nyquist noise in resistive electrical
elements \cite{Nyquist28}. In our situation, the energy loss of the
oscillator has to be counterbalanced by the detection of thermal radiation
by the antenna and the thermal current of the line. In the high temperature
limit, it leads to the usual thermodynamic ${\frac12}k_{B}T$ per degree of
freedom, with $k_{B}$ being Boltzmann constant and $T$ the radiation
temperature. Note that as a consequence, thermodynamics imposes constraints
on the scatering matrix resulting from the two principles. We will analyse
precisely these constraints in the general framework of quantum networks.

The classical result on the link between fluctuation and dissiaption was
extended to take into account the quantum statistical properties of
fluctuations\cite{Callen51,Landau84}. A general approach of these relations
was widely studied in the framework of linear response theory \cite
{Kubo66,Landau}. In particular, in the zero temperature limit, the detected
field corresponds to the vacuum fluctuations of the electromagnetic field
and the induced energy of the oscillator is the zero point energy ${\frac12}%
\hbar \omega _{0}$, with $\omega _{0}$ the resonance frequency of the
oscillator.

The elementary systems described up to now as well as more complex devices
to be studied later in this paper may be described by using a systematic
approach which may be termed as ``quantum network theory''. Initially
designed as a quantum extension of the classical theory of electrical
networks \cite{Meixner63}, this theory was mainly developed through
applications to optical systems \cite{Yurke84,Gardiner88}. It has also been
viewed as a generalized quantum extension of the linear response theory
which is of interest for electrical systems as well \cite{Courty92}. It is
fruitful for analyzing non-ideal quantum measurements containing active
elements \cite{Francesca98,Grassia99}.

\begin{figure}[htb]

\centerline{\psfig{figure=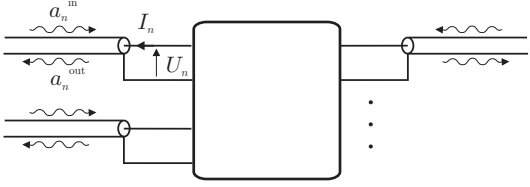,width=7cm}}
\vspace{4mm}

\caption{Representation of an electrical circuit as a quantum network. The
central box is a reactive multipole which connects noise lines corresponding
to the fluctuations entering the system, either by dissipative or by active
elements. For example, the upper left port $n$ with voltage $U_{n}$ and
current $I_{n}$ is connected to a line of impedance $R_{n}$ with inward and
outward fields $a_{n}^{{\rm in}}$ and $a_{n}^{{\rm out}}$.}
\label{Network}
\end{figure}

In this quantum network approach, the various fluctuations entering the
system, either by dissipative or by active elements, are described as input
fields in a number of lines as depicted on \ref{Network}.

For example, a resistance $R$ is modeled as a semi-infinite coaxial line $a$
with characteristic impedance $R$. In this line, current and voltage may be
treated as quantum fields propagating in a two dimensional space-time.

Free field operators $a^{{\rm in}}$ and $a^{{\rm out}}$ can be defined as
the Fourier components of $I^{{\rm in}}$and $I^{{\rm out}}$ and related to
the current and voltage at the end of the line: 
\begin{eqnarray}
I &=&\sqrt{\frac{\hbar \left| \omega \right| }{2}R}\left( a^{{\rm out}}-a^{%
{\rm in}}\right)   \nonumber \\
U &=&\sqrt{\frac{\hbar \left| \omega \right| }{2R}}\left( a^{{\rm out}}+a^{%
{\rm in}}\right)   \label{defnyquist}
\end{eqnarray}
They are normalized so that they obey the standard commutation relations 
\begin{eqnarray}
\left[ a^{{\rm in}}\left[ \omega \right] ,a^{{\rm in}}\left[ \omega ^{\prime
}\right] \right]  &=&\left[ a^{{\rm out}}\left[ \omega \right] ,a^{{\rm out}}%
\left[ \omega ^{\prime }\right] \right]   \nonumber \\
&=&2\pi \ \delta \left( \omega +\omega ^{\prime }\right) \ \varepsilon
\left( \omega \right)   \label{commutfree}
\end{eqnarray}
where $\varepsilon \left( \omega \right) $ denotes the sign of the frequency 
$\omega $. This relation just means that the positive and negative frequency
components correspond respectively to the annihilation $a_{\omega }$ and
creation $a_{\omega }^{\dagger }$ operators of quantum field theory 
\begin{equation}
a^{{\rm in}}\left[ \omega \right] =a_{\omega }\theta \left( \omega \right)
+a_{-\omega }^{\dagger }\theta \left( -\omega \right) 
\end{equation}
$\theta \left( \omega \right) $ denotes the Heavyside function. Similarly,
mechanical fluctuations as well as electromagnetical fluctuations can be
expressed in term of these normalized fields.

To characterize the fluctuations of these noncommuting operators, we use the
correlation function defined as the average value of the symmetrized
product. With stationary noise, the correlation function depends only on the
time difference 
\begin{eqnarray}
\left\langle a^{{\rm in}}\left( t\right) \cdot a^{{\rm in}}\left( t^{\prime
}\right) \right\rangle &=&\sigma _{aa}^{{\rm in}}\left( t-t^{\prime }\right)
\nonumber \\
\left\langle a^{{\rm in}}\left[ \omega \right] \cdot a^{{\rm in}}\left[
\omega ^{\prime }\right] \right\rangle &=&2\pi \ \delta \left( \omega
+\omega ^{\prime }\right) \ \sigma _{aa}^{{\rm in}}\left[ \omega \right]
\end{eqnarray}
The dot symbol denotes a symmetrized product for quantum operators.

In the case of a thermal bath, the noise spectrum is 
\begin{equation}
\sigma _{aa}^{{\rm in}}\left[ \omega \right] =\frac{1}{\exp \frac{\hbar
\left| \omega \right| }{k_{B}T_{a}}-1}+\frac{1}{2}=\frac{1}{2}\coth \frac{%
\hbar \left| \omega \right| }{2k_{B}T_{a}}
\end{equation}
One recognizes the black body spectrum or the number of bosons per mode for
a field at temperature $T_{a}$ and a term $\frac{1}{2}$ corresponding to the
quantum fluctuations. The energy per mode will be denoted in the following
as an effective temperature $\Theta _{a}$ 
\begin{equation}
k_{B}\Theta _{a}=\hbar \left| \omega \right| \sigma _{aa}^{{\rm in}}\left[
\omega \right] =\frac{\hbar \left| \omega \right| }{2}\coth \frac{\hbar
\left| \omega \right| }{2k_{B}T_{a}}
\end{equation}
In the high temperature limit the classical energy for an harmonic field of $%
k_{B}T_{a}$ per mode is recovered. In the low temperature limit, the energy $%
\frac{\hbar \left| \omega \right| }{2}$ corresponding to the ground state of
a quantum harmonic oscillator is obtained. Note that the term $\frac{1}{2}$
corresponding to the zero point quantum fluctuations was added by Planck so
that the difference with the classical result $k_{B}T_{a}$ tends to zero in
the high temperature limit \cite{Planck}.

These results are easily translated to obtain the expression of the Johnson
Nyquist noise power 
\begin{equation}
\sigma _{U_{n}U_{n}}\left[ \omega \right] =2R\hbar \left| \omega \right|
\sigma _{aa}^{{\rm in}}\left[ \omega \right] =2Rk_{B}\Theta _{a} 
\label{thermal} 
\end{equation}
Our symmetric definition of the noise power spectrum leads to a factor $2$
difference with the electronic convention where only positive frequencies
are considered.

The fields entering and leaving the network can be represented with column
vectors ${\bf a}^{{\rm in}},{\bf a}^{{\rm out}}$ with components $a_{n}^{%
{\rm in}},$and $a_{n}^{{\rm out}}$ Input fields corresponding to different
lines commute with each other. For simplicity, we also consider that the
fields entering through the various ports are uncorrelated with each other.

The whole network is then associated with a scattering ${\bf S}$ matrix,
also called repartition matrix \cite{Feldmann}, describing the
transformation from the input fields to the output ones 
\begin{equation}
{\bf a}^{{\rm out}}={\bf S\ a}^{{\rm in}}
\end{equation}
The output fields $a^{{\rm out}}$ are also free fields which obey the same
commutation relations \ref{commutfree} as the input ones. In other words, $%
{\bf S}$ matrix is unitary. More generally, the unitarity of the ${\bf S}$
matrix is required to ensure the quantum consistency of the description. In
the following section, we will make use of this property to deduce general
properties of amplifiers. For passive systems the unitarity of ${\bf S}$
matrix ensures as well the verification of the two principles of
thermodynamics. However, this unitarity is also valid in the presence of
amplificating device.

\begin{figure}[htb]

\centerline{\psfig{figure=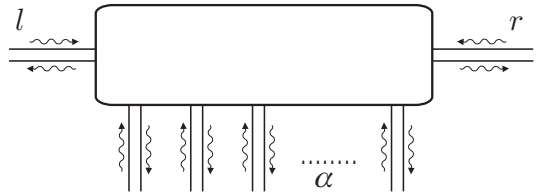,width=7cm}}
\vspace{2mm}

\caption{Quantum measurement as scattering by a network. The input signal $%
l^{in}$ enters the measurement device. An estimator is provided by the
outgoing field $r^{out}$. The other lines $\protect\alpha$ correspond to the
various active and passive elements in the measureing device.}
\label{Mesure}
\end{figure}

For the analysis of the measurement, two fields can be isolated, the input
signal $l^{{\rm in}}$ and the output detection field $r^{{\rm out}}.$ The
detection corresponds to figure \ref{Mesure}. The detection is performed
with the output detection signal $r^{{\rm out}}$. It is a linear combination
of the incoming signal $l^{{\rm in}}$ and of input fields in the various
noise lines. We normalize this expression so that the coefficient of
proportionality appearing in front of the incoming signal $l^{{\rm in}}$ is
reduced to unity. With this normalization, we obtain an estimator $\widehat{l%
}^{{\rm in}}$ which is just the sum of the true signal $l^{{\rm in}}$ to be
measured and of an equivalent input noise:

\begin{equation}
\widehat{l}^{{\rm in}}=l^{{\rm in}}+\sum_{\alpha }\mu _{\alpha }\alpha ^{%
{\rm in}}
\end{equation}
where $\alpha ^{{\rm in}}$ denote the various input fields corresponding to
the active and passive elements in the detection device. The added noise
spectrum $\Sigma _{ll}$ is obtained as 
\begin{equation}
\Sigma _{ll}=\sum_{\alpha }\left| \mu _{\alpha }\right| ^{2}\sigma _{\alpha
\alpha }^{{\rm in}}
\end{equation}

Within this approach, the succession of the different sages of the
measurement are studied by using the output field of one stage as the input
field of the next one.

\section{Measurement with active systems}

We consider in this section the amplifying stage of the measurement. Quantum
noise associated with linear amplifiers has been the subject of numerous
works. In the line of thought initiated by early works on
fluctuation-dissipation relations, active systems have been studied in the
optical domain when maser and laser amplifiers were developed \cite
{Heffner62,Haus62,Gordon63}. General thermodynamical constraints impose the
existence of fluctuations for amplification as well as dissipation
processes. The added noise determines the ultimate performance of linear
amplifiers \cite{Caves82,Loudon84} and plays a key role in the question of
optimal information transfer in optical communication systems \cite
{Gordon62,Takahasi65}.

Most practical applications of amplifiers in measurements involve ideal
operational amplifiers operating in the limits of infinite gain, infinite
input impedance and null output impedance. In order to deal with the
pathologies that could arise in such a system, we consider that it operates
with a feedback loop which fixes its effective gain and effective impedances 
\cite{Courty99}.

In the framework of quantum networks, the noise sources of the amplifier
can be modelled with two lines $a$ and $a^{\prime }$.

\begin{figure}[htb]

\centerline{\psfig{figure=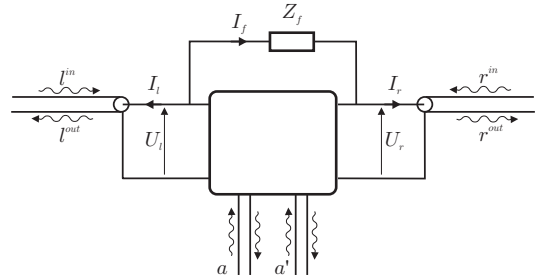,width=7cm}}
\vspace{2mm}

\caption{Measurement of an electrical signal $l^{{\rm in}}$ with an ideal
operational amplifier working in the limit of infinite gain with a reactive
feedback $Z_{f}$. The left (input) port $l$ correspond to the measured
system and the a right (output) port $r$ provides the detection signal $r^{%
{\rm out}}$. The noise sources of the amplifier are modeled as input fields
in the two noise lines $a$ and $a^{\prime}$}
\label{Ampliop}
\end{figure}

As depicted on figure \ref{Ampliop}, by coupling two coaxial lines denoted $%
l $ and $r$ respectively on the left port and the right port of the
amplifier, one realizes a model of amplifying stage. The left line comes
from a transducer so that the inward field $l^{{\rm in}}$ plays the role of
the signal to be measured. Meanwhile, the right line goes to an electrical
meter{\it \ }so that the outward field $r^{{\rm out}}$ plays the role of the
meter readout. In connection with the discussions of Quantum Non Demolition
measurements \cite{Braginsky92,Grangier92}, $l^{{\rm out}}$ appears as the
back-action field sent back to the monitored system and $r^{{\rm in}}$
represents the fluctuations coming from the readout line. A reactive
impedance $Z_{f}$ acts as feedback for the amplifier.

We now present the electrical equations associated with this measurement
device. We first write the characteristic relations between the voltages and
currents 
\begin{eqnarray}
U &=&U_{l}=U_{r}+Z_{f}I_{f}  \nonumber \\
I &=&I_{l}+I_{f}  \label{amplifier}
\end{eqnarray}
Here, $U_{p}$ and $I_{p}$ are the voltage and current at the port $p$, i.e.
at the end of the line $p=l$ or $r$, while $U$ and $I$ are the voltage and
current noise generators associated with the operational amplifier itself
(see Fig.1). $Z_{f}$ is the impedance feedback. All equations are implicitly
written in the frequency representation and the impedances are functions of
frequency. Equations (\ref{amplifier}) take a simple form because of the
limits of infinite gain, infinite input impedance and null output impedance
assumed for the ideal operational amplifier. We also suppose that the fields
incoming through the various ports are uncorrelated with each other as well
as with amplifier noises.

To obtain compact equations as well as for physical interpretations, we
introduce here a voltage noise $U$ and a current noise $I$ as linear
combination of the fields $a^{{\rm in}}$ and $a^{\prime {\rm in}}$.

\begin{eqnarray}
U\left[ \omega \right] &=&\sqrt{2\hbar \left| \omega \right| R_{a}}\left( a^{%
{\rm in}}\left[ \omega \right] -a^{\prime {\rm in}}\left[ -\omega \right]
\right)  \nonumber \\
I\left[ \omega \right] &=&\sqrt{\frac{2\hbar \left| \omega \right| }{R_{a}}}%
\left( a^{{\rm in}}\left[ \omega \right] +a^{\prime {\rm in}}\left[ -\omega %
\right] \right)  \label{deflignes}
\end{eqnarray}
Note the presence of a conjugated field $a^{\prime {\rm in}}\left[ -\omega %
\right] =a^{\prime {\rm in}}\left[ \omega \right] ^{\dagger }$ associated to
the amplification process. In the following, the conjugation of $a^{\prime }$
will be assumed.

The parameter $R_{a}$ is determined by the ratio between voltage and current
noise spectra of the amplifier 
\begin{equation}
R_{a}=\sqrt{\frac{\sigma _{UU}}{\sigma _{II}}}  \label{R0}
\end{equation}
We can deduce from (\ref{deflignes}) that the voltage and current
fluctuations $U$ and $I$ obey the following commutation relations 
\begin{eqnarray}
\left[ U\left[ \omega \right] ,U\left[ \omega ^{\prime }\right] \right] &=&%
\left[ I\left[ \omega \right] ,I\left[ \omega ^{\prime }\right] \right] =0 
\nonumber \\
\left[ U\left[ \omega \right] ,I\left[ \omega ^{\prime }\right] \right]
&=&2\pi \ \hbar \omega \ \delta \left( \omega +\omega ^{\prime }\right)
\label{UU}
\end{eqnarray}
Hence, voltage and current fluctuations verify Heisenberg inequalities which
determine the ultimate performance of the ideal operational amplifier used
as a measurement device \cite{Courty99}.

The fields $a^{{\rm in}}$ and $a^{\prime {\rm in}}$ are described by
temperatures $T_{a}$ and $T_{a^{\prime }}$. We have assumed that these
fluctuations are the same for all field quadratures, i.e. that the amplifier
noises are phase-insensitive. Although this assumption is not mandatory for
the forthcoming analysis, we also consider for simplicity that the specific
impedance $R_{a}$ is constant over the spectral domain of interest.

We use the characteristic equations (\ref{amplifier},\ref{defnyquist},\ref
{deflignes}) associated with the amplifier and the lines to write the output
fields $l^{{\rm out}}$ and $r^{{\rm out}}$ in terms of input fields $l^{{\rm %
in}}$, $r^{{\rm in}}$ and of amplifier noise sources $U$ and $I$ 
\begin{eqnarray}
l^{{\rm out}} &=&-l^{{\rm in}}+\sqrt{\frac{2}{\hbar \left| \omega \right|
R_{l}}}U  \nonumber \\
r^{{\rm out}} &=&-r^{{\rm in}}-2\frac{Z_{f}}{\sqrt{R_{r}R_{l}}}l^{{\rm in}} 
\nonumber \\
&&+\sqrt{\frac{2}{\hbar \left| \omega \right| R_{r}}}\left( \frac{R_{l}+Z_{f}%
}{R_{l}}U-Z_{f}I\right)   \label{inout}
\end{eqnarray}
The unitarity of the input output relations for $l^{{\rm out}}$ and $r^{{\rm %
out}}$ is ensured by the non commutation of the current and voltage noises.

In order to characterize the performance of the measurement device in terms
of added noise, we introduce the estimator $\widehat{l}^{{\rm in}}$ of the
signal $l^{{\rm in}}$ as it may be deduced from the knowledge of the meter
readout $r^{{\rm out}}$ 
\begin{eqnarray}
\widehat{l}^{{\rm in}} &=&-\frac{\sqrt{R_{r}R_{l}}}{2Z_{f}}r^{{\rm out}} 
\nonumber \\
&=&l^{{\rm in}}+\frac{\sqrt{R_{l}R_{r}}}{2Z_{f}}r^{{\rm in}}-\left( \frac{1}{%
Z_{f}}+\frac{1}{R_{l}}-\frac{1}{R_{0}}\right) \frac{\sqrt{R_{l}R_{0}}}{2}a^{%
{\rm in}}  \nonumber \\
&&+\left( \frac{1}{Z_{f}}+\frac{1}{R_{l}}+\frac{1}{R_{0}}\right) \frac{\sqrt{%
R_{l}R_{0}}}{2}a^{\prime {\rm in}}  \label{estim}
\end{eqnarray}
The estimator $\widehat{l}^{{\rm in}}$ would be identical to the measured
signal $l^{{\rm in}}$ in the absence of added fluctuations. Hence the noise
added by the measurement device is described by the supplementary terms
assigned respectively to the Nyquist noise $r^{{\rm in}}$ in the readout as
well as Nyquist noises $a^{{\rm in}}$ and $a^{\prime {\rm in}}$ in the two
lines representing amplification noises. Notice that proper fluctuations of $%
l^{{\rm in}}$ are included in the signal and not in the added noise.

The whole added noise is characterized by a spectrum obtained as a sum of
the uncorrelated noise spectra associated with these Nyquist noises 
\begin{eqnarray}
\Sigma  &=&\frac{R_{l}R_{r}}{4\left| Z_{f}\right| ^{2}}\sigma _{rr}^{{\rm in}%
}+\frac{R_{l}R_{0}}{4}\left| \frac{1}{Z_{f}}+\frac{1}{R_{l}}-\frac{1}{R_{0}}%
\right| ^{2}\sigma _{aa}^{{\rm in}}  \nonumber \\
&&+\frac{R_{l}R_{0}}{4}\left| \frac{1}{Z_{f}}+\frac{1}{R_{l}}+\frac{1}{R_{0}}%
\right| ^{2}\sigma _{a^{\prime }a^{\prime }}^{{\rm in}}  \label{sigma}
\end{eqnarray}
The Nyquist spectra are given by thermal equilibrium relations (\ref{thermal}%
) with temperatures $\Theta _{r}$, $\Theta _{f}$, $\Theta _{a}$. The optimal
sensitivity is obtained by matching the amplifier noise impedance $R_{0}$
with the impedance of the left line$R_{l}$.

We rewrite (\ref{estim},\ref{sigma}) under the simple forms 
\begin{eqnarray}
\widehat{l}^{{\rm in}} &=&\frac{1}{G}r^{{\rm out}}=l^{{\rm in}}+a^{\prime 
{\rm in}}-\frac{1}{G}\left( r^{{\rm in}}-a^{{\rm in}}+a^{\prime {\rm in}%
}\right)   \nonumber \\
G &=&-\frac{2Z_{f}}{\sqrt{R_{r}R_{l}}}
\end{eqnarray}
$G$ corresponds to the gain of the amplification for the normalized fields.
In the limit of very lare gain the added noise reads: 
\begin{eqnarray}
\widehat{l}^{{\rm in}} &=&l^{{\rm in}}+a^{\prime {\rm in}}  \nonumber \\
\Sigma  &=&\sigma _{a^{\prime }a^{\prime }}^{{\rm in}}  \label{SigmaMin}
\end{eqnarray}
Finally this added noise is still decreased by going to a temperature $T_{b}$
as low as possible. At the limit of a null temperature, we recover the
optimum of $3\,{\rm dB}$ added noise which is the same as for
phase-insensitive linear amplifiers \cite{Takahasi65}.

It is also possible analyse the succession of two amplifications

$r^{{\rm out}}$ is itself a quantum field it can be send at the iput of tan
other amplifier as a field $l^{^{\prime }{\rm in}}=r^{{\rm out}}\ $and can
be measured in a similar manner with a quantum operational amplifier leading
to an estimator 
\begin{equation}
\hat{r}^{{\rm out}}=r^{{\rm out}}+b^{\prime {\rm in}}-\frac{1}{G^{\prime }}%
\left( r^{\prime {\rm in}}-b^{\prime {\rm in}}+b^{{\rm in}}\right)
\end{equation}
Where the quantitities $b^{{\rm in}},b^{\prime {\rm in}},r^{\prime {\rm in}}$
and $G^{\prime }$ correspond to the second detection. As far as $r^{{\rm out}%
}$ is concerned, this is equivalet to a single detection. However, if we
consider the estimator $\widehat{l}^{{\rm in}}$ for the two successive
amplification of the signal $l^{{\rm in}}$ one obtains: is concerned, the
added noise is negligeable thanks to the large amplification of the first
detection. As a consequence, as soon as only the information on the first
system $l^{{\rm in}}$ is concerned, the second detection can be considered
as noiseless. In these condition the result $r^{{\rm out}}$ of the first
measurement can be considered as a classical quantity in the further stage
of the signal processing.

\begin{eqnarray}
\widehat{l}^{{\rm in}} &=&l^{{\rm in}}+a^{\prime {\rm in}}-\frac{1}{G}\left(
r^{{\rm in}}-a^{{\rm in}}+a^{\prime {\rm in}}+b^{\prime {\rm in}}\right)  
\nonumber \\
&&+\frac{1}{GG\prime }\left( r^{\prime {\rm in}}-b^{\prime {\rm in}}+b^{{\rm %
in}}\right) 
\end{eqnarray}
As aconsequence, with a large gain of the first amplification the only
important noise source is $a^{\prime {\rm in}}.$

\section{Analysis of a real device: the cold damped accelerometer}

We come to the discussion of the ultimate performance of the cold damped
capacitive accelerometer designed for fundamental physics experiments in
space \cite{Grassia99}.

\begin{figure}[htb]

\centerline{\psfig{figure=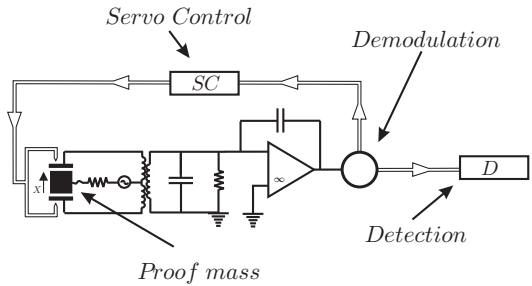,width=7cm}}
\vspace{2mm}

\caption{Scheme of the capacitive sensor. The proof mass is placed between
two electrodes formed by the inner walls of the accelerometer cage. The
position dependent capacitances are polarized by an AC sinewave source which
induces a mean current at frequency $\protect\omega _{t}$ in the symmetrical
mode. The mass displacement is read as the current induced in the
antisymmetric mode. An additional capacitance is inserted to make the
antisymmetric mode resonant with $\protect\omega _{t}$. The signal is
detected after an ideal operational amplifier with capacitive feedback
followed by a synchronous demodulation. The impedance of the detection line
plays the role of a further resistance $R_{r}$. The detected signal then
feds the servo loop used to keep the mass centered with respect to the cage.}
\label{Accelero}
\end{figure}

The central element of the capacitive accelerometer is a parallelepipedic
proof mass placed inside a box. The walls of these box are electrodes
distant from the mass off a hundred micrometers. The proof mass is kept at
the center of the cage by an electrostatic suspension. Since a three
dimensional electrostatic suspension is instable, it is necessary to use an
active suspension.

In the cage reference frame, an acceleration is transformed in an inertial
force acting on the proof mass. The force necessary to compensate this
inertial force is measured. In fact, as in most ultrasensitive measurements,
the detected signal is the error signal used to compensate the effect of the
measured phenomenon.

The essential elements of the accelerometer are presented in figure \ref
{Accelero}. The proof mass and the cage form two condensators. Any mass
motion unbalances the differential detection bridge and provides the error
signal. In order to avoid low frequency electrical noise, the electrical
circuit is polarized with an AC\ voltage with a frequency of a hundred
kilohertz. After demodulation, this signal is used for detection and as an
error signal for a servo control loop which allows to keep the mass centered
in its cage.

Furthermore, the derivative of this signal provides a force proportional to
the mass velocity and simulates a friction force. This active friction is
called cold damping since it may be noiseless. More precisely, the effective
temperature of the fluctuations of this active friction is much lower than
the physical temperature of the device.

\begin{figure}[htb]

\centerline{\psfig{figure=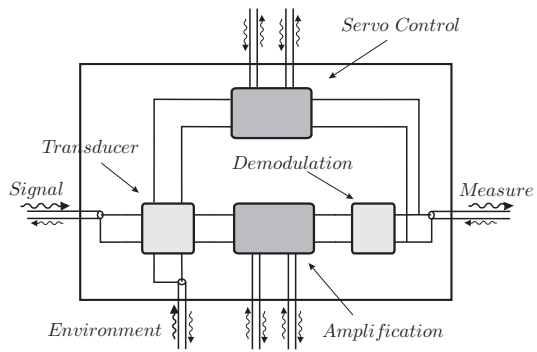,width=7cm}}
\vspace{2mm}

\caption{Scheme of the capacitive sensor represented as a quantum network.
The essential elements of the network are represented. In particular, the
input line, the mechanical dissipation corresponding to the coupling of the
transducer with the environment, the detection amplifier with its two noise
lines, the detection line and the feedback amplifier}
\label{Accelero_Network}
\end{figure}

The detection is performed with the output detection signal $r_{1}^{{\rm out}%
}$. It is a linear combination of the external force $F_{ext}$ and of input
fields in the various noise lines. We normalize this expression so that the
coefficient of proportionality appearing in front of the external force $%
F_{ext}$ is reduced to unity. With this normalization, we obtain a force
estimator $\widehat{F}_{ext}$ which is just the sum of the true force $%
F_{ext}$ to be measured and of an equivalent input force noise. In the
absence of feedback, the force estimator reads \cite{Grassia99}: 
\begin{equation}
\widehat{F}_{ext}=F_{ext}+\sum_{\alpha }\mu _{\alpha }\alpha ^{{\rm in}}
\label{estimfree}
\end{equation}
where $\alpha ^{{\rm in}}$ denote the various input fields corresponding to
the active and passive elements in the accelerometer.

When the feedback is active, the servo loop efficiently maintains the mass
at its equilibrium position and the velocity is no longer affected by the
external force $F_{ext}$. The residual motion is interpreted as the
difference between the real velocity of the mass and the velocity measured
by the sensor. This means that the servo loop efficiently corrects the
motion of the mass except for the sensing error. However the sensitivity to
external force is still present in the correction signal. Quite remarkably,
in the limit of an infinite loop gain and with the same approximations as
above, the expression of the force estimator $\widehat{F}_{ext}$ is the same
as in the free case \cite{Grassia99}.

The added noise spectrum $\Sigma _{FF}$ is obtained as 
\begin{equation}
\Sigma _{FF}=\sum_{\alpha }\left| \mu _{\alpha }\right| ^{2}\sigma _{\alpha
\alpha }^{{\rm in}}
\end{equation}
We have evaluated the whole noise spectrum $\Sigma _{FF}$ for the specific
case of the instrument proposed for the $\mu $SCOPE space mission devoted to
the test of the equivalence principle. Some of the main parameters of this
system are listed below 
\begin{eqnarray}
M=0.27{\rm \ kg} &\qquad &H_{m}=1.3\times 10^{-5}{\rm \ kg\ s}^{-1} 
\nonumber \\
\frac{\Omega }{2\pi }\simeq 5\times 10^{-4}{\rm \ Hz} &\qquad &\frac{\omega
_{t}}{2\pi }\simeq 10^{5}{\rm \ Hz}  \nonumber \\
R_{a}=0.15\times 10^{6}\ \Omega &\qquad &\Theta _{a}=1.5{\rm \ K}
\label{parameters}
\end{eqnarray}
$M$ is the mass of the proof mass, $H_{m}$ is the residual mechanical
damping force, $\frac{\Omega }{2\pi }$ is the frequency of the measured
mechanical motion, $\frac{\omega _{t}}{2\pi }$ is the operating frequency of
the electrical detection circuit. $R_{a}$ and $\Theta _{a}$ are the
characteristic impedance and temperature of the amplifier.

In these conditions, the added noise spectrum is dominated by the mechanical
Langevin forces 
\begin{eqnarray}
\Sigma _{FF} &=&2H_{m}k_{B}\Theta _{m}  \nonumber \\
&=&1.1\times 10^{-25}\left( {\rm kg\ m\ s^{-2}}\right) ^{2}/{\rm Hz}
\end{eqnarray}
This corresponds to a sensitivity in acceleration 
\begin{equation}
\frac{\sqrt{\Sigma _{FF}}}{M}=1.2\times 10^{-12}{\rm \ m\ s^{-2}}/\sqrt{{\rm %
Hz}}
\end{equation}
Taking into account the integration time of the experiment, this leads to
the expected instrument performance corresponding to a test accuracy of $%
10^{-15}$.

In the present state-of-the-art instrument, the sensitivity is thus limited
by the residual mechanical Langevin forces. The latter are due to the
damping processes in the gold wire used to keep the proof mass at zero
voltage \cite{Willemenot97}. With such a configuration, the detection noise
is not a limiting factor. This is a remarkable result in a situation where
the effective damping induced through the servo loop is much more efficient
than the passive mechanical damping. This confirms the considerable interest
of the cold damping technique for high sensitivity measurement devices.

Future fundamental physics missions in space will require even better
sensitivities. To this aim, the wire will be removed and the charge of the
test mass will be controlled by other means, for example UV photoemission.
The mechanical Langevin noise will no longer be a limitation so that the
analysis of the ultimate detection noise will become crucial for the
optimization of the instrument performance. This also means that the
electromechanical design configuration will have to be reoptimized taking
into account the various noise sources associated with detection \cite
{Grassia99}.

\section{Conclusion}

In this analysis of chained quantum measurements, the preamplification
settles the Von Neumann problem of determining where the measurement result
can be considered as classical.

Notice that in our analysis of the cold damped accelerometer, this fact is
important to treat properly the feedback. The result of the measurement is
used to act on the quantum system. This action is provided by a quantum
signal, however, we have shown that the noise added in the active feedback
loop is negligeable compared to the noise of the first stage.

\end{document}